\documentclass[aps,twocolumn]{revtex4}
\usepackage{graphicx}
\newcommand{\lint}{\int_{-\Lambda}^\Lambda}
\newcommand{\trho}{\widetilde\rho}

\begin{document}


\title{Competing density-wave orders in a\\ one-dimensional hard-boson model}

\author{Paul Fendley}

\affiliation{Department of Physics, University of Virginia,
Charlottesville, VA 22904-4714}

\author{K. Sengupta and Subir Sachdev}
\affiliation{Department of Physics, Yale University, P.O. Box
208120, New Haven CT 06520-8120}

\date{\today}

\begin{abstract}
We describe the zero-temperature phase diagram of a model of
bosons, occupying sites of a linear chain, which obey a
hard-exclusion constraint: any two nearest-neighbor sites may have at
most one boson. A special case of our model was recently proposed
as a description of a ``tilted'' Mott insulator of atoms trapped
in an optical lattice. Our quantum Hamiltonian is shown to
generate the transfer matrix of Baxter's hard-square model. Aided
by exact solutions of a number of special cases, and by numerical
studies, we obtain a phase diagram containing states with
long-range density-wave order with period 2 and period 3, and also
a floating incommensurate phase. Critical theories for the various
quantum phase transitions are presented. As a byproduct, we show
how to compute the Luttinger parameter in integrable theories with
hard-exclusion constraints.
\end{abstract}

\maketitle

\section{Introduction}
\label{sec:intro} In recent years, the study of quantum models
with multiple competing ground states has emerged as an important
theme in the study of strongly-correlated many-body quantum
systems. For example, in the cuprates it is clear that states with
density-wave order at a variety of wavevectors play an important
role in the physics at low carrier concentrations.

In this paper, we introduce a simple one-dimensional quantum model
which displays a multiplicity of ground states. Despite its
simplicity, it exhibits ({\em i\/}) gapped states with
commensurate density-wave order (with periods of 2 and 3 lattice
spacings), ({\em ii\/}) gapless regions with ``floating''
incommensurate, quasi-long-range density-wave correlations, and
({\em iii\/}) gapped states which preserve translational symmetry.
A special case of our model appeared in a recent study \cite{ssg}
of atoms trapped in optical lattices \cite{bloch}, and so an
experimental study of the phase diagram presented here may be
feasible. More generally, we offer our model as a simple
laboratory, with many exactly solvable cases, for the interplay of
density-wave orders with multiple periods in quantum systems.

Our model is expressed in terms of the bose operator $d_j$, which
annihilates a boson on site $j$, and the boson number operator
\begin{equation}
n_j \equiv d^{\dagger}_j d_j . \label{nj}
\end{equation}
The ``hard'' boson condition allows no more than 1 boson on any
pair of nearest-neighbor sites, and hence all states obey the
constraints
\begin{equation}
n_j \leq 1~~~;~~~n_j n_{j+1} = 0. \label{constraints}
\end{equation}
In the study of Mott insulators in optical lattices \cite{ssg},
the $d_j$ boson represents a {\em dipole} excitation, consisting
of a particle-hole pair bound on nearest neighbor sites of the
optical lattice. This microscopic dipole interpretation will not
be crucial to our analysis here, and so we will refer to $d_j$
simply as a boson.

The boson Hamiltonian we study is
\begin{equation}
\mathcal{H} = \sum_j \left[ -w \left( d_j + d^{\dagger}_j \right)
+ U n_j + V n_j n_{j+2} \right] \label{h}
\label{ham}
\end{equation}
Note that the total number of bosons is not conserved, and it is
possible to create and annihilate bosons out of the vacuum. This
is natural in the dipole interpretation of the boson, as a
particle-hole pair can be created or annihilated from the
background Mott insulator. There is also no explicit boson hopping
term; as was shown in Ref.~\onlinecite{ssg}, boson hopping is
generated by the combination of the constraints in
(\ref{constraints}) and single-site terms already in
$\mathcal{H}$, and so it is not necessary to include an explicit
hopping. $U$ is a chemical potential for the bosons, while $V$
is a ``nearest''-neighbor interaction, ``nearest'' meaning two
sites apart, the closest two bosons can come.
One can of course rescale out one of the couplings
to obtain a two-parameter Hamiltonian, but it will be convenient to
keep all three.
The case $V=0$ of $\mathcal{H}$ was studied in
Ref.~\onlinecite{ssg}; we have also streamlined the earlier
notation of the coupling constants to a form suitable for our
analysis here.
Without the constraints
(\ref{constraints}), the Hamiltonian would be trivially
solvable. With them, its analysis becomes quite intricate.

The ground state of our Hamiltonian (\ref{ham}) can exhibit several
kinds of order, depending on the couplings. The Hamiltonian will favor
having bosons on every other site if we have an attractive
``nearest''-neighbor interaction, or a chemical potential favoring the
creation of particles.  We will show that this leads to a regime of
Ising-type order, with a spontaneously broken ${\bf Z}_2$ symmetry,
translation by one site.  If the chemical potential still favors
creating particles, but there is a repulsive ``nearest'' neighbor
potential, then the ground state will favor having a particle on every
third site. This sort of order breaks a ${\bf Z_3}$ symmetry,
translation by one or two sites.  When the the two kinds of ordered
states have nearly the same energies, we will show that there exists
an incommensurate phase. In the incommensurate phase, bosons appear on
every other or every third site.

The one-dimensional quantum Hamiltonian (\ref{ham}) looks
unusual because of the single-particle creation and annihilation operators.
However, it in fact has already appeared in a very different context:
it arises from taking the (Euclidean) time continuum limit of a
two-dimensional classical statistical-mechanical model, the hard-square model
\cite{square}. This model describes
the statistical mechanics of square tiles placed
on the sites of a square lattice. Each square tile is rotated by
45$^\circ$ from the principal axes of the square lattice, and the
area of each tile is twice the area of a single plaquette of the
lattice. Tiles are not allowed to overlap, so this means
nearest-neighbor sites cannot both be occupied: putting a tile on
every other site of the square lattice covers all of space.

It is easy to see that the Hilbert space of our one-dimensional
quantum theory is identical to the space of states along a line of the
hard-square model. In the quantum theory, the Hilbert space consists
of bosons which are restricted one to a site and forbidden to be
nearest neighbors.  In the two-dimensional classical theory, the
squares appear at most one to a site, and are forbidden to be nearest
neighbors. It is less obvious that one can obtain the quantum
Hamiltonian (\ref{ham}) by taking a limit of the classical transfer
matrix. We show in the Appendix how to do this, by
taking the (Euclidean) time direction to be along the diagonals of the
square lattice. Roughly speaking, the chemical potential $U$ and
interaction $V$ correspond to a chemical potential and interaction
strength for the squares. The precise relations are derived in the Appendix.

The phase diagram of the hard-square model has been studied in
Ref.~\onlinecite{husesquare}. We will explain in section II how these
results apply to the ground state of our quantum Hamiltonian. An
extremely useful result is that the hard-square model is integrable
for some values of the Boltzmann weights \cite{square}.  In terms of
the Hamiltonian (\ref{ham}), this amounts to two lines in our
two-parameter space of couplings:
\begin{equation}
 w^2=UV+ V^2 .
\label{intlines}
\end{equation}
There is a single critical point along each of these lines; they are
in the universality classes of the Ising tricritical point and of the
three-state Potts model critical point. We will show how these
critical points describe transitions to and from the ${\bf Z}_2$ and
${\bf Z}_3$ ordered phases.

In section \ref{sec:phase} we discuss the different regions of phase
space. In section \ref{sec:potts} we discuss in depth the phase with
order of period 3, the phase with spontaneously-broken ${\bf Z}_3$
symmetry. We present numerical results to support the analytic
arguments.  In sections \ref{sec:float} and \ref{sec:bethe}, we
discuss the incommensurate ``floating'' phase in depth. In section
\ref{sec:float} we derive an effective Hamiltonian for this region,
and prove that an incommensurate phase exists by bosonizing it. In
section \ref{sec:bethe} we calculate the size of this phase by using
the Bethe ansatz to compute the effective Luttinger parameter
exactly. This enables us to find the region in which vortex-type
perturbations are irrelevant and do not destroy the
incommensurability.  Unfortunately, this region occupies a fairly
small region of parameter space, so it is not possible to see this
result numerically.

\section{Phase diagram}
\label{sec:phase}

\begin{figure}
\centerline{\includegraphics[width=3.2in]{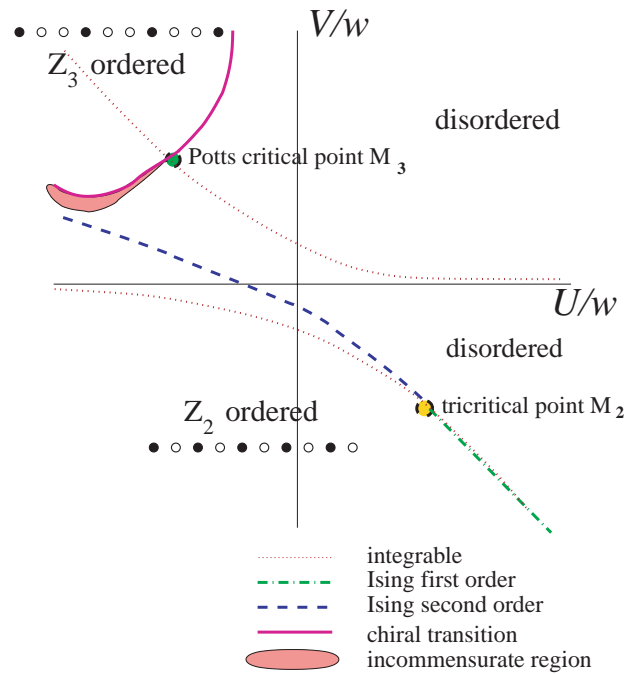}}
 \caption{Ground-state phase diagram of $\mathcal{H}$
as a function of $U/w$ and $V/w$. Schematic representations of the
ground states with density waves of period 2 and 3 are shown. The
dotted lines indicate the positions of the integrable lines in
(\protect\ref{intlines}). The critical points along these lines
(\protect\ref{so},\protect\ref{crit}) are labeled $M_{2,3}$. The
extent of the incommensurate region has been greatly exaggerated
for clarity: it is drawn to scale in Fig~\protect\ref{phase2}.
There could also be an incommensurate region adjacent to the
chiral transition line about $M_3$, but we have no direct evidence
for this possibility.} \label{phase}
\end{figure}

Our main results are summarized in the phase diagram shown in
Fig~\ref{phase}.
We discuss the phases and phase transitions in this figure by
considering a number of limiting cases.

\subsection{${\bf Z}_2$ order}
\label{sec:ising}
First we consider the case $V=0$, which was studied earlier in
Ref.~\onlinecite{ssg}. Although the Hamiltonian looks simple in this limit,
the prohibition of nearest-neighbor bosons means that
this case is not solvable except as $U/w\to \pm\infty$.
For large positive $U/w$ we obtain a
featureless ground state with a small density of bosons and which
breaks no translational symmetry, as is easily seen from
$\mathcal{H}$. This is a disordered, or liquid, phase.
Similarly, for $U/w$ large and negative, the energy
is minimized by states with a maximal density of bosons: there are
2 degenerate states of this type, each with every second site
occupied; in other words, there is a density wave of period 2, and
translational symmetry is spontaneously broken. It was shown numerically
in Ref.~\onlinecite{ssg} that there is a second-order Ising critical
point which separates these phases at $U/w   = - 1.308 \ldots$.

One can also find the location of the transition from Ising order
to disorder easily in the limit
$V \rightarrow -\infty$ and $U\approx -V$.
With $|U|$ and $|V|$ both large, we can ignore the
$w$ term, and then $\mathcal{H}$ becomes a model of classical
particles. The empty state with no particles has energy 0, while
the maximally occupied states with density waves of period 2 have
energy per lattice site of $(U+V)/2$. There is therefore a
first-order transition between these states at $U=-V$. Actually,
it is quite easy to find the position of this
first-order transition to order $(w/V)^2$. Using standard
second-order perturbation theory in $w$, the energy per site of
the empty state is $-w^2/U$. Similarly, the energy per site of the
maximally occupied state is $(U+V)/2 + w^2 /(2(U+2V))$. Equating
these two energies we estimate the position of the first-order
transition to be
\begin{equation}
\frac{U}{V} = -1+ \frac{w^2}{V^2} \label{fo}
\end{equation}

The integrability of the model along the lines (\ref{intlines}) lets
us understand how this transition occurs at finite $V$ and $w$. In
fact, the condition (\ref{fo}) coincides precisely with one of
the integrable lines in (\ref{intlines}). Indeed, the arguments of
Huse \cite{husesquare} and the exact results \cite{square,baxpearce}
imply that integrable line is precisely along the
co-existence line of these two phases, and hence the first-order
boundary is given exactly by (\ref{fo}), with no corrections at
higher order in $w/V$.

At the value
\begin{equation}
\frac{V}{w}=-\left(\frac{\sqrt{5}+1}{2}\right)^{5/2}
\label{so}
\end{equation}
there is a critical point along the integrable line. Thus the
first-order transition line must terminate.  Huse argued that this
integrable quantum critical point was a tricritical point
separating a first-order boundary from a line of second-order
Ising transitions. Indeed, the computation of exact critical
exponents at this point \cite{square,baxpearce} shows that this
critical point in the hard-square model is in the same
universality class as the tricritical point in the Ising model.
This tricritical point in the continuum limit is described by the
minimal conformal field theory with central charge $c=7/10$
\cite{huse}.  Therefore, the line of first-order transitions
determined by (\ref{fo}) is present for $V/w < -
((\sqrt{5}+1)/2)^{5/2}$, where it terminates at the Ising
tricritical point $M_2$ shown in Fig~\ref{phase}.  There is a line
of second-order Ising transitions on the other side of $M_2$, but
this is {\em not} on the integrable line (\ref{intlines}); this
second-order line clearly passes through the point (\ref{so})
discussed above.  The exact computations show that the integrable
line for $0>V/w >- ((\sqrt{5}+1)/2)^{5/2}$ is in the ${\bf Z}_2$
ordered phase \cite{square, baxpearce}. This is consistent with
the finite value (\ref{so}) of $U_c$ for $V=0$; at $V=0$ the
integrable line is at $U=-\infty$.

\subsection{${\bf Z}_3$ order}
\label{sec:vpotts}

Consider the case $V \rightarrow \infty$, $|U|\ll V$. The large
positive value of $V$ now prohibits bosons separated by 2 lattice
spacings, and hence we effectively have the constraint
\begin{equation}
n_j n_{j+2} = 0 \label{pconst}
\end{equation}
along with the constraints (\ref{constraints}). The physics in
this region can be understood using reasoning parallel to that in
Section~\ref{sec:ising}. As before, for large positive $U/w$, we
obtain a featureless ground state with a small density of bosons,
which breaks no lattice symmetry. For $U/w$ large and
negative, the ground state has a maximal density of bosons: with
the constraints (\ref{constraints}) {\em and} (\ref{pconst}) there
are now 3 degenerate states of this type, each with every third
site occupied. Here we thus have a density wave of period 3.

To study this phase with spontaneously-broken ${\bf Z}_3$ symmetry
at finite $V$, we can use results for the other integrable line,
where $V/w>0$ in (\ref{fo}).
Along this line there is a quantum critical point at \cite{square}
\begin{equation}
\frac{V}{w}  = \left( \frac{\sqrt{5}+1}{2} \right)^{5/2}.
\label{crit}
\end{equation}
This critical point is in the universality class of the three-state
Potts model \cite{square,baxpearce}. In the continuum limit this is
described by the minimal conformal field theory with central charge
$4/5$ \cite{huse}. For $w/V$ smaller than this value and along this line, the
exact results indeed show that the ${\bf Z}_3$ translation symmetry is
broken. For $w/V$ larger than this value and along this line, there is
no order. Thus along the integrable line, the transition is in the
universality class of the usual order-disorder transition in the
three-state Potts model; similar ${\bf Z}_3$ quantum
criticality occurs in an XXZ chain in a staggered field \cite{z3}.
However, as opposed to the Ising case, the
transition away from this integrable line is not
of the usual type \cite{husesquare}: it is related that of the
chiral clock model \cite{ostlund}.
We will discuss the nature of the transition between these states as a
function of $U/w$ in Section \ref{sec:potts}, including the
possibility of additional intermediate states with incommensurate
periods.

\subsection{Competing orders}
\label{sec:deg}

The most complicated and interesting limit is
$V \rightarrow \infty$, $U \approx - 3V$. where there
is competition between density-wave orders of period
2 and period 3. Initially, as in Section~\ref{sec:ising}, we
anticipate that with $|V/w|$ and $|U/w|$ both large, we can neglect
quantum effects and consider a model of classical particles. The
period-2 and period-3 density-wave states then have energies per site
$(U+V)/2$ and $U/3$ respectively, so that the
ground states become degenerate at $U=-3V$.
In fact, at $U=-3V$ in this limit, there are many
degenerate states: any
state where successive particles are separated by either 2 or 3
lattice spacings is within the ground state manifold. This results
in an extensive classical entropy at $U=-3V$. This
extensive degeneracy is lifted only by quantum effects, and
perturbation theory in $w$ is therefore highly non-trivial. The
analysis of quantum effects in this regime will be presented in
Sections \ref{sec:float} and \ref{sec:bethe}.
We will establish that there is an
intermediate gapless phase with incommensurate spin correlations.
We will show as well that
the gapped state with no broken symmetry also intervenes between
the period-2 density-wave state and the incommensurate phase, as
shown in Fig.~\ref{phase}.

We will argue in Section \ref{sec:potts} that there also should be
an incommensurate phase near the Potts critical point $M_3$. The
simplest and most plausible possibility is that this phase and the
one for $-U\approx 3V \gg 1$ are the same. We have drawn the phase
diagram in this fashion.

\section{Density-wave order of period 3}
\label{sec:potts}

The density-wave state of period 3 appears when $V$ is large and
positive, so that bosons with separations of both 1 and 2 lattice
spacings are suppressed. This section and the next will discuss
the manner in which quantum fluctuations destroy this period-3
order.

It is useful to begin with the limit $V \rightarrow \infty$, noted
above in Section~\ref{sec:vpotts}. At first glance, there appear
to be only two distinct possibilities for the ground states: a
featureless, low density state at large positive $U/w$, and a
density-wave state of period 3 at large negative $-U/w$. We might
also anticipate that these states are separated by a second-order
critical point in the universality class of the 3-state Potts
model. However, the situation is more subtle \cite{husesquare}.

It is useful to introduce order parameters, $\Psi_p$, which
characterize density-wave orders of period $p$ (in general, $p$
need not be an integer):
\begin{equation}
\Psi_p = \sum_j n_j e^{i 2 \pi j/p}
\label{ppotts}
\end{equation}
For $p=2$, the Ising order parameter $\Psi_2$ is real. However,
for the $p=3$ case we consider in this section, the Potts order
parameter $\Psi_3$ is complex. Clearly, the state with density
wave order of period 3 has $\langle \Psi_3 \rangle \neq 0$, while
the featureless, low density state has $\langle \Psi_3 \rangle =
0$.

Using the usual symmetry criteria, we can write down a continuum
quantum field theory describe the onset of Potts order. The action
should be invariant under translations, under which $\Psi_3
\rightarrow e^{i 2 \pi \ell/3} \Psi_3$, with $\ell$ integer. A
crucial point, noted by Huse and Fisher \cite{husefisher} in a
different context, is the behavior under spatial inversion:
\begin{equation}
x \rightarrow - x~~~~,~~~~\Psi_3 \rightarrow \Psi_3^{\ast}
\label{inversion}
\end{equation}
where $x$ is the continuum spatial co-ordinate. Finally, we note
that $\Psi_3$ is invariant under time reversal. These symmetry
constraints lead to the following proposed effective action for the quantum
field theory ($\tau$ is imaginary time):
\begin{eqnarray}
\mathcal{S}_3 &=& \int dx d\tau \left[ i \alpha \Psi_3^{\ast}
\partial_x \Psi_3 + \mbox{c.c} + |\partial_\tau \Psi_3 |^2 + v^2 |\partial_x \Psi_3
|^2\right. \nonumber \\
&+& \left. r |\Psi_3 |^2 + v \Psi_3^3 + \mbox{c.c.} + \ldots
\right] \label{schiral}
\end{eqnarray}
The critical point associated with this field theory occurs at $\alpha=r=0$,
and is in the same universality class as the critical point in the
three-state Potts model. This is described by a conformal field theory of
central charge $4/5$, and the dimensions of all the operators are known.

The exact results show that the integrable critical point $M_3$ indeed
is in the universality class of the three-state Potts model
\cite{square,baxpearce}.  It separates a state with density-wave order
of period 3 from the gapped translationally-invariant state
\cite{huse,square}.  Just because the critical point is in the same
universality class as the three-state Potts model does not mean that
all the physics in the region of the critical point is the same. The
key difference is the presence here of the linear spatial derivative
proportional to $\alpha$, which is clearly allowed under the
symmetries (\ref{inversion}). This term is {\em chiral}, in that it
breaks rotational symmetry.  The exact results show no evidence of
chiral behavior, so the effective theory describing the integrable
line presumably corresponds to $\alpha=0$. Varying $r$ at $\alpha=0$
therefore moves one along the integrable line, and
describes the usual physics of the three-state Potts model: a
second-order transition between ${\bf Z}_3$ order and disorder.

When one moves off the integrable line (taking $\alpha\ne 0$ in the
effective theory), the theory in the universality class of the {\em
chiral clock model} \cite{ostlund}.  The chiral perturbation,
$\alpha$, is known to be relevant at the ordinary Potts critical point
\cite{cardy}, with scaling dimension 1/5; the perturbing operator has
dimensions $9/5$ and breaks Lorentz invariance.
Thus for $\alpha\ne 0$,
the quantum phase transition associated
with the vanishing of density-wave order of period 3 is in general
{\em not} in
the same universality class as the usual order/disorder transition in
the ordinary 3-state Potts model. The integrable point $M_3$ is an
isolated multicritical point;
all other points on the phase boundary of the period-3 state
must have $\alpha \neq 0$, and display the physics of the chiral
clock model.
Note the contrast with the
corresponding transition associated with density-wave order of period
2: this transition is in the Ising universality class because $\Psi_2$
is real, and then the term proportional to $\alpha$ is a vanishing
total derivative.

The chiral clock model is generally believed to exhibit an
incommensurate phase \cite{ostlund,husefisher,cardy}. In the region
near the multicritical point $M_3$, this occurs when the chiral
perturbation ($\alpha\ne 0$) dominates over the thermal perturbation
($r \ne 0$). This region is therefore quite narrow, because $\alpha$,
of dimension $1/5$, is much less relevant than $r$, which is of dimension
$6/5$. Thus $r$ must effectively vanish for the effects of the chiral
perturbation to be felt. In the
next section, we will show that the incommensurate phase does have a
finite (albeit small) width far from the multicritical point, where
$|U|$ and $V$ are large, with $U=\approx V/3$.  There is no indication
that any more new physics intervenes in between this region and the
incommensurate region near $M_3$, so we presume that these two phases
are the same, as we have indicated in Fig.~\ref{phase}.

In the remainder of this section we will describe our numerical
study of the physics in the limit $V \rightarrow \infty$, which
corresponds to being on the portion of the phase boundary of the
period-3 state {\em above} the multicritical point $M_3$ in
Fig~\ref{phase}. Our numerical results here are inconclusive, as
we see little deviation from the pure 3-state Potts physics for
the system sizes examined. We will study the portion of the phase
boundary {\em below} $M_3$ in Section~\ref{sec:float}, and there
we shall definitively establish the existence of a gapless
incommensurate phase, driven by the presence of the chiral
perturbation; the ordering wavevector, $K = 2 \pi/p$, of this
incommensurate phase obeys $K > 2 \pi /3$. A simple fluctuation
analysis of $\mathcal{S}_3$ indicates that $K - 2 \pi/3$ is
proportional to $\alpha$. As $\alpha$ vanishes at $M_3$ where $K=
2 \pi /3$, these results suggest the conjecture that there is a
small intermediate, gapless, incommensurate phase also above the
point $M_3$, but with a change in sign of $\alpha$ leading to an
ordering wavevector $K < 2 \pi /3$. Our numerical studies below,
however, do not show any specific evidence in support of this
conjecture. This is possibly because the scaling dimension of the
chiral perturbation is quite small, and very large system sizes
are needed before its effects are perceptible. Furthermore, we
shall find in the study of the portion of the phase boundary below
$M_3$ in Section~\ref{sec:float} that the region of incommensurate
spin correlations is extremely small, and it is likely that a
similar feature holds in the above $M_3$ region being studied
here.

The numerical analysis for large $V$ is simplified by simply
eliminating all states with an energy determined by $V$. This is
equivalent to extending the constraints (\ref{constraints}) to
\begin{equation}
n_j \leq 1~~~;~~~n_j n_{j+1} = 0~~~;~~~n_j n_{n+2} = 0,
\label{pconstraints}
\end{equation}
and working with the simplified Hamiltonian
\begin{equation}
\mathcal{H}_s = \sum_j \left[ -w \left( d_j + d^{\dagger}_j
\right) + U n_j  \right] \label{hps}
\end{equation}
The exact diagonalization of $\mathcal{H}_p$ can then be carried
out for system sizes $N \le 21$.

\begin{figure}
\centerline{\includegraphics[width=3.2in]{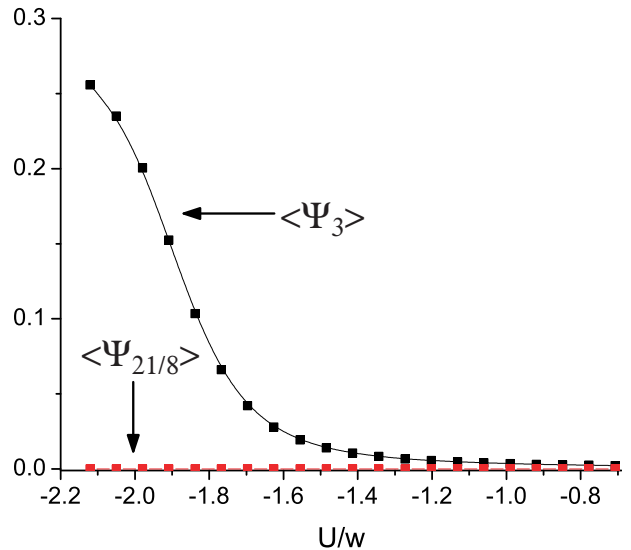}}
\caption{Density-wave order, $\langle \Psi_p \rangle$, induced by
an external potential at wavevector $2 \pi/p$ and magnitude
$10^{-4}$. There appears to be onset of spontaneous density-wave
order with period $p$ at the smaller values of $U/w$.}
\label{susplot1}
\end{figure}

\begin{figure}
\centerline{\includegraphics[width=3.2in]{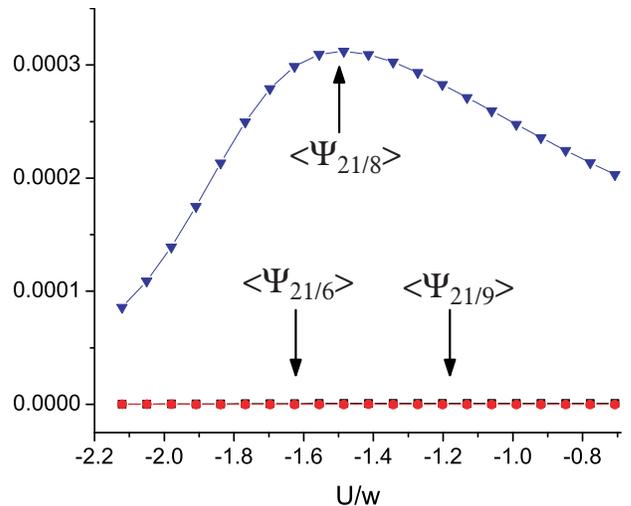}}
\caption{As in Fig.~\protect\ref{susplot1}, but with a greatly
expanded vertical scale. All of the periods shown above show no
sign of spontaneous order.} \label{susplot2}
\end{figure}

First, we searched for the preferred values of $p$ in a system of
size $N=21$. We applied a small external cosine potential on the
bosons of magnitude $10^{-4}$ and wavevector $2\pi/p$, and
measured the resulting values of $\langle \Psi_p \rangle$ as a
function of $U/w$. The results are shown in Figs~\ref{susplot1}
and~\ref{susplot2}. Fig.~\ref{susplot1} shows that value of
$\Psi_3$ is much larger than the response at another nearby
wavevector.
Figure~\ref{susplot2} shows similar data, but with a greatly
expanded vertical scale.
The response at periods $p \neq 3$ is always {\em much} smaller,
and so we see no sign of spontaneous order at any other
wavevectors. Interestingly, the largest secondary response is at
the wavevector $16 \pi/21$. This is contrary to our expectations
earlier based on the influence of the chiral perturbation in
(\ref{schiral}), where we anticipated incommensurate order at
wavevectors smaller than $2 \pi /3$. However, given the very small
susceptibility, and the small system sizes considered we are not
able to interpret any of this data in support of an incommensurate
phase. Any such phase, if present, must be over an extremely small
parameter regime, as is found in Section~\ref{sec:float}

Ignoring the presence of a possible intermediate incommensurate
phase, we now see if our data can be interpreted in the context of
a direct transition from the gapped density-wave state of period
3, to a gapped ``liquid'' state which does not break any
translational symmetries. We expect such a transition to be in the
Potts universality class. Indeed, we now show that our data are in
complete accord with such an assumption, with excellent agreement
to the expected values of the critical exponents. This is again
consistent with an as yet undetected influence of the chiral
perturbation in the present analysis (the chiral perturbation is
more clearly detected in Section~\ref{sec:float}).

\begin{figure}
\centerline{\includegraphics[width=3.2in]{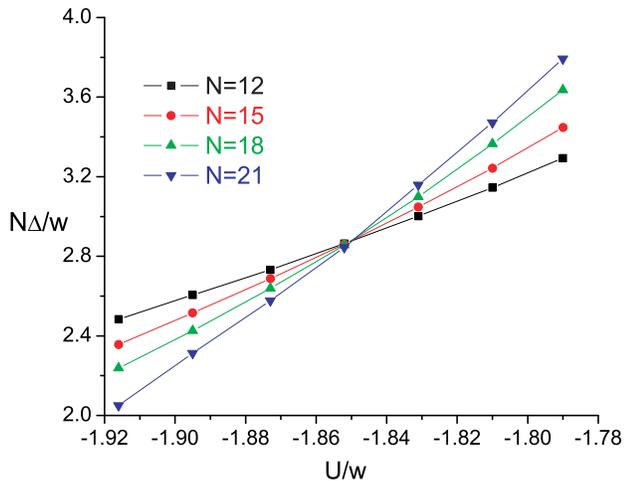}} \caption{Plot
for the energy gap $\Delta$ as a function of $U$ for different
system sizes. The data for different system sizes cross at
$U_c/w=-1.852$.} \label{sca1}
\end{figure}

\begin{figure}
\centerline{\includegraphics[width=3.2in]{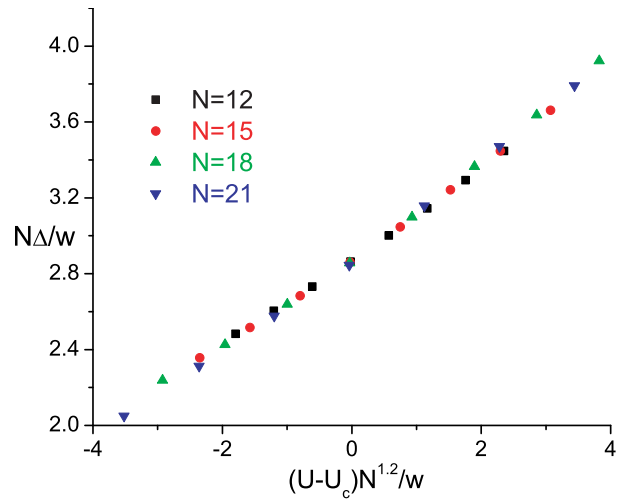}}
\caption{Scaling plot for the energy gap $\Delta$. The data for
different system sizes collapse for $\nu=0.83$ and
$U_c/w=-1.852$.} \label{sca2}
\end{figure}

First, let us determine the position of a presumed critical point
from the period-3 (Potts) density-wave state to a disordered state
using finite size scaling analysis. Near the quantum critical
point, the energy gap $\Delta$ is expected to obey the scaling
relation
\begin{eqnarray}
\Delta &=& N^{-z} \Phi \left ( N^{1/\nu}(U-U_c)\right) \label{sc1}
\end{eqnarray}
where $\Phi$ is an universal scaling function, $z$ is the dynamic
critical exponent, and $\nu$ is the correlation length exponent.
Fig.\ \ref{sca1} shows a plot of $N\Delta/w$ vs $U/w$ which
exhibits a clear crossing point at
\begin{equation}
\frac{U_c}{w}=-1.852.
\end{equation}
From Eq.\ \ref{sc1}, we thus conclude that $z=1$. Fig\ \ref{sca2}
shows a plot of $ N\Delta/w$ vs $N^{1.2}(U-U_c)/w$. We find that
the data for all $N$ collapses for $U_c/w=-1.852$, giving
$\nu=0.833$. The numerical values of $\nu$ and $z$ obtained here
are in excellent agreement with the known analytical values $z=1$
and $\nu=5/6$ for the three-state Potts model \cite{Wu82}.

\begin{figure}
\centerline{\includegraphics[width=3.2in]{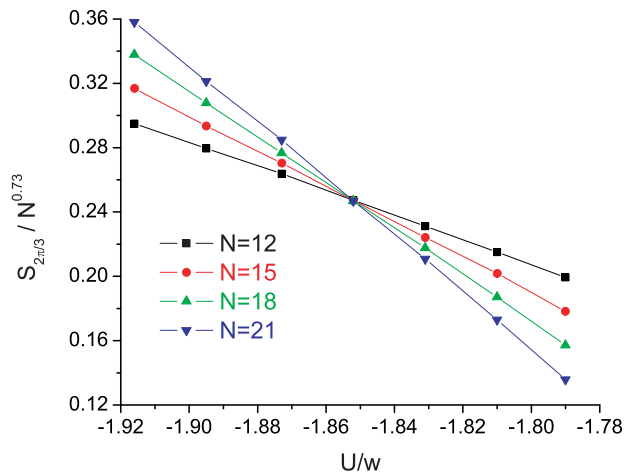}} \caption{Plot
of the equal-time structure factor $S_{2\pi/3}$ for different
system sizes. The data for different system sizes cross at
$U_c/w=-1.852$ for $\eta=0.27$} \label{sca3}
\end{figure}

As a final test of the three-state Potts universality class, we
calculate the equal-time structure factor of the Potts order
parameter $\Psi_3$ (Eq.\ \ref{ppotts}) given by
\begin{eqnarray}
S_{2\pi/3} &=& \frac{1}{N} \left\langle \Psi_3^{\ast} \Psi_3
\right\rangle.
 \label{sc2}
\end{eqnarray}
The structure factor is expected to scale as $N^{2-z-\eta}$ near
the critical point. A plot of $S_{2\pi/3}/N^{0.73}$ vs $U/w$, as
shown in Fig.\ \ref{sca3}, shows excellent crossing at
$U_c/w=-1.852$. From this, knowing $z=1$, we identify
$\eta=0.27$, which is again in good agreement with the known exact
value $\eta=4/15$ \cite{Wu82}.

\section{Competition between density-wave orders of periods 2 and
3: the incommensurate state}
\label{sec:float}

We noted in Section~\ref{sec:deg} that in the classical limit
($w=0$), the period-2 and period-3 states become degenerate for
$U=-3V$. Here we shall study the quantum physics near this
degeneracy point. As will become clear, the physics is quite
complicated, but we are able to obtain a complete picture by a
combination of perturbation theory, bosonization, and Bethe ansatz
methods. Our primary result is that the gapped period-2 and period
3 are separated by 2 intermediate phases, and all these phases are
separated by 3 second-order quantum phase transitions (see
Fig.~\ref{phase}). Above the period-2 phase is the gapped,
translationally invariant state, accessed by a conventional Ising
transition. Above this is a gapless, incommensurate phase reached
across a Kosterlitz-Thouless transition. Finally, we reach the
gapped period-3 phase via a Pokrovsky-Talapov transition
\cite{pt}.

We will establish these results by a careful study of the region
of the phase diagram where
\begin{equation}
0 < w, |U+3V| \ll |U|, V \label{e1}
\end{equation}
It will become clear from our analysis that the non-trivial region
of the phase diagram is where
\begin{equation}
|U + 3V| \sim \frac{w^2}{V}. \label{e2}
\end{equation}
So we introduce the new dimensionless coupling $\sigma$, and
parameterize
\begin{equation}
U \equiv -3V \left[ 1 + \left(\frac{w}{V} \right)^2 \sigma
\right]. \label{e3}
\end{equation}
The above conditions on the interesting regime can now be restated
as follows
\begin{equation}
0 < w \ll V~~~;~~~\sigma \sim 1 \label{e4}
\end{equation}
We shall follow the physics as $\sigma$ is tuned from large
negative to large positive values, while maintaining (\ref{e1}).

\subsection{The effective Hamiltonian}

\begin{figure}
\centerline{\includegraphics[width=2.5in]{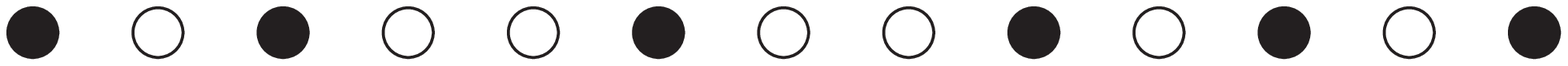}} \caption{The
state $|\ldots 23322 \ldots \rangle$. Filled circles are $d_j$
bosons, and empty circles have no bosons.} \label{state}
\end{figure}

As we mentioned in Section~\ref{sec:deg}, under the condition
(\ref{e4}), the important states are those in which successive
$d_j$ bosons are separated by either 2 or 3 lattice spacings. We
denote such a state by the sequence of lattice spacings {\em e.g.}
$| \ldots 23323233 \ldots \rangle$; see Fig.~\ref{state}.
We wish to develop an operator formalism for describing such
states. We shall introduce two distinct formalisms below, but it
should always be kept in mind that both describe identical Hilbert
spaces with identical spectra.

In the first approach, we describe the states in terms of
`defects' or `domain walls', or `kinks' in the ordered period-2 density-wave
states. This description focuses on the 3's in a background of
2's. We view each 3 as a new boson, created by the operator
$t^{\dagger}_\ell$, located at the center of the 3 defect. Note
that the $\ell$ sites lie at the midpoints of the links of the $j$
sites of the original model $\mathcal{H}$. Thus the state $|
\ldots 2223222 \ldots \rangle$ represents a single $t$ boson. An
arbitrary state in the low-energy
subspace can be described by a
configuration of $t$ bosons. The resulting $t$ boson states obey
some constraints: there can only be at most one $t$
boson per site, and successive $t$ bosons must be separated by
$3,5,7,9,11\ldots$ sites.

Second-order perturbation theory
also allows us to obtain the effective Hamiltonian for the $t$
bosons to order $(w/V)^2$. It is useful to first tabulate the
energies of different configurations of the original $d$ bosons in
different environments, to order $(w/V)^2$. For a $d$ boson
between 22 bonds we have
$$
E_{22} = U + V + \frac{w^2}{U + 2 V}.
$$
Similarly for a $d$ boson between 33 bonds we have
$$
E_{33} = U + \frac{w^2}{U}.
$$
Finally for a $d$ boson between 23 bonds we have
$$
E_{23} = U + \frac{V}{2} + \frac{w^2}{U+V}.
$$
From these expressions, we can obtain the energy of a $t$ boson
well-separated from all other $t$ bosons:
\begin{eqnarray*}
E_t &=& 2 E_{23} - \frac{5}{2} E_{22} \nonumber \\
&=& V \left[ \frac{3(1+\sigma)}{2} \left( \frac{w}{V} \right)^2 +
\mathcal{O} \left(\frac{w}{V} \right)^3 \right]
\end{eqnarray*}
Similarly, we obtain the interaction energy of two adjacent $t$
bosons
\begin{eqnarray*}
E_{t, {\rm int}} &=& 2 E_{23} + E_{33} - 4 E_{22} - 2 E_t
\nonumber \\ &=& V\left[ -\frac{1}{3} \left( \frac{w}{V} \right)^2
+ \mathcal{O} \left(\frac{w}{V} \right)^3 \right]
\end{eqnarray*}
Finally, moving on to the $t$ boson hopping amplitude, a similar
perturbation theory shows that it is $w^2 /(U+V)$. Collecting all
these results, we can now write down the effective Hamiltonian for
the $t$ bosons, valid in the regime (\ref{e4}).  On a system of
$N$ sites the Hamiltonian $\mathcal{H}$ in (\ref{h}) is effectively
\begin{eqnarray}
\frac{\mathcal{H}_t}{V} &=& -\frac{N}{2}\left(2 + (1+3 \sigma)
\left( \frac{w}{V} \right)^2 \right) \nonumber \\ &+& \left(
\frac{w}{V} \right)^2 \sum_\ell \left[ -\frac{1}{2} \left(
t_{\ell+2}^{\dagger} t_{\ell} + t_{\ell}^{\dagger} t_{\ell+2}
\right) \right.  \\
&+& \left. \frac{3}{2} (1+ \sigma) t_\ell^{\dagger} t_\ell -
\frac{1}{3} t^{\dagger}_{\ell+3} t_{\ell+3} t_\ell^{\dagger}
t_\ell \right] + \mathcal{O} \left(\frac{w}{V} \right)^3 \nonumber
\label{ht}
\end{eqnarray}
We reiterate that $\mathcal{H}_t$ must be solved under the constraints
that there can only be at most one $t$ boson per site, and successive
$t$ bosons must be separated by $3,5,7,9,11\ldots$ sites. To this order in
$w/V$ the Hamiltonian conserves the number of $t$ bosons.
Except for
the latter constraint, this Hamiltonian is very much like that of an
XXZ magnet. The hopping term in (\ref{ht}) is akin to spin exchange,
the chemical potential is like an effective magnetic field $h_{eff} =
3(1+\sigma)/2$, and the interaction is akin to the $S_z S_z$
interaction between neighboring sites. The analogous (ferromagnetic)
XXZ model has
anisotropy $J_z/J_x=1/3$, so it is in a gapless regime. We shall show in
sect.\ \ref{sec:bethe} that this model is as well, for appropriate
values of the effective magnetic field.

Before turning to analysis of $\mathcal{H}_t$, it is useful to
introduce a complementary analysis of the physics valid under the
conditions (\ref{e4}). Now, we consider defects or kinks
between ordered states with period 3. So we treat the 2's as
bosons moving in background 3's. Such bosons are created by the
operator $p_j$, and note that this boson resides on the sites of
the original lattice. The energy of an isolated $p$ boson is
\begin{eqnarray*}
E_p &=& 2 E_{23} - \frac{5}{3} E_{33} \nonumber \\
&=& V \left[- \left(\sigma +\frac{4}{9}\right) \left( \frac{w}{V} \right)^2 +
\mathcal{O} \left(\frac{w}{V} \right)^3 \right]
\end{eqnarray*}
Also, the interaction energy of two adjacent $p$ bosons is
\begin{eqnarray*}
E_{p,{\rm int}} &=& 4 E_{23} + E_{22} - 4 E_{33} - 3 E_p \nonumber
\\
&=& V \left[ -\frac{1}{3} \left( \frac{w}{V} \right)^2 +
\mathcal{O} \left(\frac{w}{V} \right)^3 \right]
\end{eqnarray*}
Finally, the hopping matrix element of the $p$ bosons is the same
as that of the $t$ bosons. Collecting these results yields the
following effective description of $\mathcal{H}$ in the regime
(\ref{e4}):
\begin{eqnarray}
\frac{\mathcal{H}_p}{V} &=& -\frac{N}{3}\left(3 + \frac{(1+9
\sigma)}{3} \left( \frac{w}{V} \right)^2 \right) \nonumber \\ &+&
\left( \frac{w}{V} \right)^2 \sum_j \left[ -\frac{1}{2} \left(
p_{j+3}^{\dagger} p_{j} + p_{j}^{\dagger} p_{j+3}
\right) \right.  \\
&-& \left. \left(\sigma +\frac{4}{9}\right) p_j^{\dagger} p_j - \frac{1}{3}
p^{\dagger}_{j+2} p_{j+2} p_j^{\dagger} p_j \right] + \mathcal{O}
\left(\frac{w}{V} \right)^3. \nonumber \label{hp}
\label{hamp}
\end{eqnarray}
The constraints on the $p$ bosons are that each site can have at
most one $p$ boson, and successive $p$ bosons can only be
separated by intervals of $2,5,8,11,14,\ldots$. Note that except
for the constraints, this description
also resembles a (ferromagnetic) XXZ model with
the same anisotropy $\Delta=J_z/J_x=1/3$.

It is important to note that the Hilbert space and spectra of
$\mathcal{H}_t$ in (\ref{ht}) and $\mathcal{H}_p$ in (\ref{hp})
should be identical, although this is by no means obvious. Some
simple consistency checks can however easily be performed: the
vacuum state of $\mathcal{H}_t$ with no $t$ bosons should have the
same energy as the state of $\mathcal{H}_p$ with the maximal
number of $p$ bosons. The reader can easily check that this
requirement, and its converse, do indeed hold. More generally, we
can easily see from the manner in which we defined the states that
if a state has $N_t$ $t$ bosons under $\mathcal{H}_t$, then the
same state has $N_p$ $p$ bosons under $\mathcal{H}_p$ where
\begin{equation}
2 N_p + 3 N_t = N, \label{e6}
\end{equation}
and $N$ is assumed to be a multiple of 6.

\begin{figure}
\centerline{\includegraphics[width=3.5in]{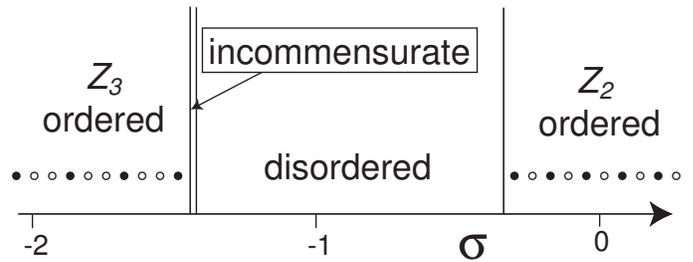}}
\caption{Ground states of $\mathcal{H}$, $\mathcal{H}_t$, or
$\mathcal{H}_p$ as a function of $\sigma$, under the conditions
(\protect\ref{e4}). The dimensionless coupling $\sigma$ is defined
in (\protect\ref{e3}); so the horizontal axis above corresponds to
moving downward along the extreme left of Fig~\protect\ref{phase}.
The transition from the $Z_2$ ordered state to the disordered
state at $\sigma=-1/3$ is in the Ising universality class, and has
dynamic exponent $z=1$. The $Z_3$ ordered state first undergoes a
$z=2$ transition to an incommensurate phase in the
Pokrovsky-Talapov universality class at $\sigma = -13/9$, which is
then followed by a $z=1$ Kosterlitz-Thouless transition at $\sigma
= -1.422...$ (this last number is determined from the Bethe ansatz
analysis in Section~\protect\ref{sec:bethe}).} \label{phase2}
\end{figure}

Some simple considerations now allow us to deduce some important
properties of $\mathcal{H}_t$ and $\mathcal{H}_p$. Consider first
the case $\sigma \gg 1$. A glance at $\mathcal{H}_t$ shows that
each $t$ boson costs a large positive energy, and so the ground
state is the $t$ boson vacuum (which also happens to be the state
with the maximal number of $p$ bosons). This vacuum state is
clearly the state with density state of period 2. The lowest
excited state consists of a $t$ boson excitation (a kink),
at zero momentum, which has an energy $(w^2 / V)(1 + 3 \sigma)/2$
above the ground state. So we conclude that the ground state
remains a density wave of period 2 provided
\begin{equation}
\sigma > - \frac{1}{3}~~~\rightarrow \mbox{period-2 density-wave
order}
\end{equation}
This conclusion is illustrated in Fig.~\ref{phase2}, which displays
the final phase diagram emerging from the analysis of this
section.

A similar argument can be applied to determine the stability of
the period-3 density-wave state. From $\mathcal{H}_p$, this is
stable for $\sigma \ll -1$. The energy of a single $p$ boson
excited state at zero momentum is $-(w^2 / V)(13 + 9 \sigma)/2$,
and so the ground state remains a density wave of period 2
provided
\begin{equation}
\sigma < - \frac{13}{9}~~~\rightarrow \mbox{period-3 density-wave
order}
\end{equation}
This is also shown in Fig.~\ref{phase2}.

\subsection{Stability of the incommensurate phase}
\label{sec:bos}

It now remains to understand the physics for $-13/9 < \sigma <
-1/3$. The above considerations suggest that this intermediate
region has a finite density of both $t$ and $p$ bosons obeying
(\ref{e6}). If this is a compressible phase, then we expect the
ground state to be a Luttinger liquid, and the corresponding
correlations of the density of the underlying $d$ bosons to be
incommensurate. Furthermore, the transition $\sigma = -13/9$
involving the onset of a non-zero density of the $p$ bosons would
be a transition of the Pokrovsky-Talapov type \cite{pt}, with
dynamic critical exponent $z=2$ \cite{sss}, and an effective free
fermion description near the critical point. Similar
considerations can be expected to apply to the transition at
$\sigma = -1/3$, associated with the onset of a finite density of
$t$ bosons.

However, the validity of such a conclusion requires the absence of
additional perturbations, at higher order in $w/V$, which could
disrupt the stability of the Luttinger liquid phase. Of particular
importance here is the possible generation of  `vortex operators'
\cite{haldane,schulz,villain} which violate the conservation of the total
number of $t$ or $p$ bosons obeyed by the terms so far included in
$\mathcal{H}_{t,p}$. A simple analysis of the processes at higher
orders in $w/V$ in terms of the $b$ bosons shows that such
processes do indeed appear: the domain walls in the period-2
density-wave state, the $t$ bosons, can be annihilated or created
in pairs. Similarly for the period-3 state, there are processes in
which 3 $p$ bosons can be annihilated or created. We will examine
the consequences of such processes by a
combination of bosonization and Bethe ansatz methods. We will find
that the vortex perturbations are indeed irrelevant at the
$\sigma=-13/9$ critical point, so that there is a direct $z=2$
transition from the gapped period-3 density-wave state, to a
stable, gapless incommensurate phase. However, the vortex
perturbations are found to be relevant at the $\sigma=-1/3$
critical point, implying that the gapless incommensurate phase
does not extend all the way up to the period-2 density-wave state
(see Fig.~\ref{phase2}).

Here we study the stability of the Luttinger liquid phase of
$\mathcal{H}_{t,p}$ postulated above within the framework of a
long-wavelength bosonization analysis. This will allow us to place
general conditions on the Luttinger parameters required for stability.
This analysis will allow us to determine the values of the Luttinger
parameters in some important limiting cases, and show there indeed
exists an incommensurate phase stable to perturbation. The analysis
does not allow us to determine the Luttinger parameter in general
--- this we will do in the following section by a Bethe ansatz
analysis.

First, let us carry out the analysis using $\mathcal{H}_t$, and
let us assume the value of $\sigma$ is such that we are in a
compressible Luttinger liquid region. We follow the notation of
Ref.~\onlinecite{book}, and introduce the continuum field
$\theta_t$ with the action
\begin{equation}
\mathcal{S}_t = \frac{K_t}{2 \pi v} \int dx d\tau \left[
(\partial_\tau \theta_t)^2 + v^2 (\partial_x \theta_t)^2 \right]
\label{st}
\end{equation}
Here $v$ is a velocity and $K_t$ is a Luttinger parameter, normalized
so that free fermions have free fermions have $K=1$. In
terms of this field
\begin{equation}
t \sim e^{-i \theta_t}. \label{thetat}
\end{equation}
Now the vortex operator $V \sim t^2 \sim e^{-2 i \theta_t}$ is
seen to have scaling dimension $1/K_t$ under the action
$\mathcal{S}_t$, and hence the vortices are irrelevant, and the
compressible phase is stable, as long as
\begin{equation}
K_t < \frac{1}{2} .\label{iv}
\end{equation}
We will determine the value of $K_t$ as a function of $\sigma$ in
the following subsection, but here we note a simple where $K_t$
can be determined exactly. This is the limit where the density of
$t$ bosons becomes vanishingly small as we approach the period-2
density-wave state with $\sigma \nearrow -1/3$. In this low
density limit near a $z=2$ quantum transitions, it can be shown
that the $t$ particles behave like free fermions \cite{sss}, and
so we must have $K_t=1$. The same argument can also be used to
determine $v$ and so we have the important result
\begin{equation}
v = (w^2 /V) \sqrt{-1-3\sigma}~~~;~~~K_t = 1~~~\mbox{as $\sigma
\nearrow -1/3$}.
\end{equation}
Notice that $K_t$ does not obey (\ref{iv}), and hence vortices are
always relevant at the boundary of the period-2 state, as we
claimed above. As these vortices correspond to creation of pairs
of the free fermions, the resulting fermion model is easily seen
to be equivalent to the fermionized Ising model. This implies that
we have an Ising transition at $\sigma=-1/3$ directly from the
gapped period 2 ordered state to a disordered gapped state that
does not break any symmetries, as shown in Fig~\ref{phase2}.

Next, a complementary bosonization analysis can be applied to
$\mathcal{H}_p$. We express the low energy physics using a
Luttinger field $\theta_p$ with the action
\begin{equation}
\mathcal{S}_p = \frac{K_p}{2 \pi v} \int dx d\tau \left[
(\partial_\tau \theta_p)^2 + v^2 (\partial_x \theta_p)^2 \right]
\label{sp}
\end{equation}
As $\mathcal{H}_t$ and $\mathcal{H}_p$ have the same spectrum, the
velocity $v$ should be the same as that in (\ref{st}). However,
the Luttinger parameter $K_p$ is now different because the
interpretation of the field $\theta_p$ differs from
(\ref{thetat}):
\begin{equation}
p \sim e^{-i \theta_p} \label{thetap}
\end{equation}
Now the vortex operator $V^{\dagger} \sim p^3$ has scaling
dimension $9/(4K_p)$, and hence the vortices are irrelevant, and
the compressible phase is stable, as long as
\begin{equation}
K_p < \frac{9}{8} \label{ip}
\end{equation}
Again, determination of the values of $K_p$ and $v$ requires a
Bethe ansatz analysis, but exact results can be obtained in the
limit $\sigma \searrow -13/9$, where the density of $p$ bosons
becomes vanishingly small. The same arguments as applied above for
the $t$ bosons now show that
\begin{equation}
v = (w^2/V) \sqrt{2(13+9 \sigma)/9}~~~;~~~K_p = 1~~~\mbox{as
$\sigma \searrow -13/9$.}
\end{equation}
Now $K_p$ does obey the stability condition (\ref{ip}), and hence
vortices are always irrelevant at the boundary of the transition
to the period-3 density-wave state. So, as claimed earlier, we
have established the existence of a gapless, incommensurate state
in this region. At some intermediate value of $\sigma$, the vortices
will become relevant and will drive a Kosterlitz-Thouless
transition to a gapped, disordered phase.

We conclude this subsection by noting that the relationship
(\ref{e6}) implies a simple, general relationship between the
Luttinger actions in (\ref{st}) and (\ref{sp}). Indeed, the
standard connection between particle density and the $\theta$
field in bosonization theory \cite{book}, combined with
(\ref{e6}), leads to the condition
\begin{equation}
\frac{\theta_p}{\theta_t} = -\frac{2}{3}. \label{K3}
\end{equation}
This result is also consistent with our expressions for the vortex
operator $V$ noted below (\ref{thetat}) and (\ref{thetap}). Using
(\ref{K3}) in (\ref{st}) and (\ref{sp}), we also obtain the exact
relationship
\begin{equation}
\frac{K_t}{K_p} = \frac{4}{9}, \label{K1}
\end{equation}
which shows, not surprisingly, that the conditions (\ref{iv}) and
(\ref{ip}) are the same.

\section{The size of the incommensurate phase}
\label{sec:bethe}

To establish the boundaries of the incommensurate phase, we will
show that the effective Hamiltonian (\ref{ht}) (or equivalently,
(\ref{hamp})) can be solved using the Bethe ansatz. We will then
use this to determine the Luttinger parameter anywhere in the
incommensurate phase. The computation is a generalization of that
done in \cite{haldanelutt} to cases with a ``hard-core''
constraint. Bethe ansatz analyses have been done in several
related cases, the XXZ model with a hard core (not allowing down
spins to be adjacent)\cite{AB}, or a model of hard-core fermions
arising in a supersymmetric chain \cite{FSN}. The details of the
computation of the Luttinger parameter have not been presented,
however, so we describe the calculation in generality here.

\subsection{The Bethe ansatz}

The $t$ boson in the effective Hamiltonian (\ref{ht}) is a kink which
separates the two density-wave states.
We denote the Ising-ordered state with no kinks
as  $|0\rangle$, so that an eigenstate of the Hamiltonian (\ref{hamt})
with $N_t\le N/3$ kinks is
\begin{equation}
\phi^{(N_t)} = \sum_{\{i_j\}} \varphi(i_1,i_2,\dots i_f)
t^\dagger_{i_1}t^\dagger_{i_2}\dots t^\dagger_{i_f}|0\rangle
\label{phif}
\end{equation}
The $i_j$ are ordered so that $1\le i_1<i_2<\dots$, and moreover
we require that $i_{j+1} -i_j=3,5,7,\dots$; the Hamiltonian preserves
this restriction.
To simplify the analysis
in this section, we shift away the constant term in ${\cal H}_t$,
multiply by a constant,
and define the effective chemical potential $h_t=-3(1+\sigma)/2$. This yields
\begin{equation}
H = -\sum_\ell \left[ \frac{1}{2} \left(
t_{\ell+2}^{\dagger} t_{\ell} + t_{\ell}^{\dagger} t_{\ell+2}
\right) + h_t t_\ell^{\dagger} t_\ell +
\frac{1}{3} t^{\dagger}_{\ell+3} t_{\ell+3} t_\ell^{\dagger}
t_\ell \right].
\label{hamt}
\end{equation}
This Hamiltonian has the same eigenvectors as the original.
For simplicity, we also assume periodic boundary conditions.

Bethe's ansatz is that the state $\phi^{({N_t})}$ is of the form
\begin{equation}\varphi(i_1,i_2,\dots i_{N_t}) = \sum_P
A_P\,  \mu_{P1}^{i_1}\mu_{P2}^{i_2}\dots \mu_{P{N_t}}^{i_{N_t}}.
\label{phidef}
\end{equation}
for some numbers $\mu_j$ and  $A_P$, where $j=1,\dots, N_t$ and
$P \equiv (P1,P2,\dots, PN_t)$ is a permutation
of the integers $(1,2,\dots,N_t)$.
With periodic boundary conditions,
we can construct eigenstates of the translation operator $T$, which
sends $T t^\dagger_{i_1} t^\dagger_{i_2}\dots
= t^\dagger_{i_1+1} t^\dagger_{i_2+1}\dots$.
A state obeying Bethe's ansatz is an eigenstate of $T$ if
the amplitudes $A_P$ are cyclically related as
\begin{equation}
 A_{P{N_t},P1,P2,\dots P({N_t}-1)} = \mu_{P{N_t}}^N
 A_{P1,P2,\dots P{N_t}}.
\label{Acyclic}
\end{equation}
If we define the bare momenta $k_j$ via $\mu_j \equiv e^{-ik_j}$,
the eigenvalue $e^{ik_{\rm tot}}$ of $T$ is then given by
\begin{equation}
k_{\rm tot} = -i\ln\left[\prod_{j=1}^{N_t} (\mu_j)^{-1}\right] =
\sum_{j=1}^{N_t} k_j \label{ev}
\end{equation}
The relation (\ref{Acyclic}) then can be thought of as the quantization
of the momentum of a particle in a box.

The $\mu_i$ and the amplitudes $A_P$
are found by demanding that $\phi^{({N_t})}$ be an
eigenstate. It is simplest to illustrate this in the case of two kinks.
The Bethe ansatz is that
\begin{eqnarray*}
\varphi(i_1,i_2)&=&\sum_P A_P\  \mu_{P1}^{i_1}\mu_{P2}^{i_2}\\
&=&A_{12} \mu_{1}^{i_1}\mu_{2}^{i_2}
+A_{21} \mu_{2}^{i_1}\mu_{1}^{i_2}
\end{eqnarray*}
Requiring $\phi^{(2)}$ be an eigenstate of $T$ means that
$$A_{12}=A_{21} \mu_1^N,$$
with $e^{-ik_{\rm tot}}=\mu_1\mu_2$.
Operating with the Hamiltonian (\ref{hamt})
on $\phi^{(2)}$ yields
$$H\phi^{(2)}=E_t\phi^{(2)} +
\sum_{i=1}^N X_i t^\dagger_{i} t^\dagger_{i+3}$$
where
$$E_t= -
h_t N_t - \frac{1}{2}\left[\mu_1^2 + \mu_2^2 + (\mu_1)^{-2} + (\mu_2)^{-2}\right].$$
and
$$X_i = -\frac{2}{3}\varphi(i,i+3) +\varphi(i,i+1) +\varphi(i+2,i+3).$$
The Bethe equations are derived by requiring that
all the $X_i$ vanish, so that $\phi^{(2)}$ is
an eigenstate. With the Bethe ansatz, this means that
$$X_i=-\sum_P A_P\, \mu_{P1}^{i}\mu_{P2}^{i}
\left(2\Delta\mu_{P2}^3 - \mu_{P2}^{} - \mu_{P1}^{2}\mu_{P2}^3\right)=0$$
where $\Delta=1/3$.
Each of these vanishes if
$$\frac{A_{21}}{A_{12}}=-\frac{\mu_2}{\mu_1}
\frac{\mu_{2}^2\mu_1^2 +1 -2\Delta \mu_{2}^2}
 {\mu_{2}^2\mu_1^2 +1 - 2\Delta \mu_{1}^2}.$$
Combining this with the periodic boundary conditions (\ref{Acyclic}) yields
$$ \frac{A_{12}}{A_{21}}=\mu_1^N = -\frac{\mu_1(\mu_1^2\mu_2^2 +1 -\mu_1^2)}
{\mu_2(\mu_1^2\mu_2^2 +1 -\mu_2^2)}.$$ Let $\nu_j = \mu_j^2$. Then
using the fact that $e^{-ik_{\rm tot}} = \mu_1\mu_2$ gives the
Bethe equation
$$
\nu_m^{(N-2)/2} e^{-ik_{\rm tot}} = - \frac{\nu_m\nu_j +1
-\nu_m} {\nu_m\nu_j +1 -\nu_j},$$ which holds for $(m,j)=(1,2)$
and $(2,1)$. One solves these two equations for $\nu_1$ and
$\nu_2$ subject to the constraint $e^{-2ik_{\rm tot}} =
\nu_1\nu_2$. Then the corresponding eigenstate is found (up to an
overall constant) by substituting the values of $\nu_i$ into the
equation for $A_{21}/A_{12}$.

This computation can be generalized to any number of kinks ${N_t}$.
The constraint that $\phi^{({N_t})}$ be an eigenstate is basically
the same as the two-kink case: one requires that
\begin{eqnarray*}
0&=&2\Delta\varphi(...,i,i+2, ...) - \varphi(..., i,i+1, ...) \\
&&\qquad -\varphi(...,i+1,i+2,...)
\end{eqnarray*}
for any choice of $i_1,i_2,\dots i_{N_t}$ and $i$.
The trick to make this vanish is to consider the permutation
$P'$, which differs from $P$ only in that $Pm$ and $P(m+1)$ are reversed,
i.e.\ $P'=P1P2\dots P(m-1) P(m+1) Pm P(m+2)\dots P{N_t}$.
Therefore $\phi^{({N_t})}$ is an eigenstate if for all $P$ and $m$
\begin{equation}
\frac{A_{P'}}{A_P} = g(\nu_{P(m+1)},\nu_{Pm})
\label{AA}
\end{equation}
with
$$g(a,b)\equiv - \frac{\sqrt{a}(ab+1 -2\Delta a)}{\sqrt{b}
(ab+1 - 2\Delta b)}.$$
One can think of $g$ as the bare S-matrix describing
the phase shift when two kinks are interchanged.
The fact that $g(a,a)=-1$  means
that the wavefunction vanishes if two of the bare momenta $k_j$ are identical.
To find the $\nu_j$, we impose the boundary condition (\ref{Acyclic}).
Note that
\begin{eqnarray*}
\frac{A_{PN_t,P1\dots P({N_t}-1)}}
{A_{P1,P2,\dots P{N_t}}} &=&
\frac{A_{P1,PN_t,P2\dots P({N_t}-1)}}
{A_{P1,P2\dots P{N_t}}} g(\nu_{P{N_t}},\nu_{P1})\\
&=& \prod_{j=1}^{{N_t}} g(\nu_{P{N_t}},\nu_{Pj})
\end{eqnarray*}
The condition (\ref{Acyclic}) must hold for all $PN_t$, and hence all $j$.
Putting this all together yields the Bethe equations for the $\mu_j$ or
$\nu_j$:
$$\nu_j ^{N/2} = \prod_{m=1}^{N_t} g(\nu_j,\nu_m).$$
This is thus a coupled set of polynomial equations for the $\mu_i$.
Using the explicit form of $g$, and the expression (\ref{ev}) for
the translation eigenvalue $e^{ip}$, the Bethe equations simplify to
\begin{equation}
\nu_j^{(N-{N_t})/2} e^{-ik_{\rm tot}} = (-1)^{{N_t}-1}
\prod_{m=1}^{N_t} \frac{\nu_j\nu_m + 1 -2\Delta\nu_j} {\nu_j\nu_m
+1 -2\Delta\nu_m}. \label{bethe}
\end{equation}
These are a set of ${N_t}$ coupled polynomial equations for the
$\nu_j=e^{2ik_j}$, $j=1\dots {N_t}$. Solving these for a set of
$k_j$, one then finds the corresponding
eigenstate (up to an overall normalization) by using (\ref{AA}).
This eigenstate has energy
\begin{equation}
E_t= -h_tN_t -  \sum_{j=1}^{N_t} \cos(2k_j) \label{energy}
\end{equation}
These are precisely the Bethe equations and energy one obtains for
the XXZ model at $\Delta=J_z/J_x=1/3$ with $(N-{N_t})/2$ sites, if
one imposes twisted boundary conditions resulting in the factor of
$e^{-ik_{\rm tot}}$ in (\ref{bethe}). The twist and the different
number of sites result from the excluded-volume effect (the
restriction that $t$ bosons must be $3,5,7,\dots$ sites apart),
and the fact that $t$ bosons hop two sites instead of one.

In the next subsection we will use these Bethe equations to find
the size of the incommensurate phase.
Since we saw by using bosonization that the incommensurate region
is closer to the ${\bf Z_3}$-ordered phase than to the ${\bf Z_2}$-ordered
phase, the density of $p$ bosons in the incommensurate phase will be much
smaller than that of the $t$ bosons. It is thus useful to derive the
Bethe equations in the $p$ boson basis. The analogous wavefunction
$\widetilde\varphi$ must obey
\begin{eqnarray*}
0 &=&2\Delta\widetilde\varphi(...,i,i+2,...)
-\widetilde\varphi(...,i,i-1,...)\\
&&\qquad - \widetilde\varphi(...,i+3,i+2,...)
\end{eqnarray*}
with, crucially, the same $\Delta=1/3$ as for the $t$ bosons.
Going through a virtually-identical calculation yields
$$e^{3ik_{\rm tot}}=\prod_{j=1}^{N_p} \omega_j$$
\begin{equation}
E_p= -h_pN_p -
\frac{1}{2}\sum_{j=1}^{N_p} \left[\omega_j + \omega_j^{-1}\right]
\label{Ep}
\end{equation}
\begin{equation}
\omega_j^{(N+{N_p})/3} e^{ik_{\rm tot}} = (-1)^{{N_p}-1}
\prod_{m=1}^{N_p} \frac{\omega_j\omega_m + 1 -2\Delta\omega_j}
{\omega_j\omega_m +1 -2\Delta\omega_m}. \label{bethe2}
\end{equation}
where the effective chemical potential here is $h_p = \sigma + 4/9$.  Note
that the right-hand-side of the Bethe equations here are identical to
those for the $t$ bosons (and for the corresponding XXZ model); only
the left-hand-side changes.

\subsection{Calculation of the Luttinger parameter}

Since a great deal of discussion
of the size of the incommensurate phase has appeared in the literature
(see e.g.\ \cite{haldane,schulz,villain}), we feel it is worth
discussing in detail a case where the answer can be determined exactly
via the Bethe ansatz.
In this subsection we therefore derive explicitly the size of the
incommensurate phase by finding the range of $\sigma$ where the
vortex operators are irrelevant. We do this by computing the
Luttinger parameters $K_p= 9K_t/4$ of the effective Hamiltonian.
The dimension of
operators can be computed directly by using either the asymptotics
of the correlator \cite{BIK} or finite-size effects \cite{WET},
but the result is identical. In this subsection, we will
generalize Haldane's computation of the Luttinger parameter
\cite{haldanelutt} to the case at hand. The computation
we do here is more complicated because the exponent on the
left-hand-side of (\ref{bethe}) or (\ref{bethe2}) depends on the
number of kinks in the system. This is a consequence of the
restrictions on the locations of the particles: more particles
effectively reduces the size of the system.

The XXZ spin chain at $\Delta=1/3$ is in a gapless phase.  The vortex
operators are forbidden from occuring because of the $U(1)$ symmetry
of the spin chain; if one breaks this symmetry by allowing
e.g.\ $J_x\ne J_y$, then the resulting XYZ chain is indeed gapped.  One
can show using the above Bethe ansatz computation that Hamiltonians
(\ref{ht}) and (\ref{hamp}) are also in a gapless phase: the slight
difference in Bethe equations does not change this result. This
difference, however, does complicate the computation of the dimension
of the vortex operators. It also means that as opposed to the XYZ
model, there is a gapless phase with broken $U(1)$ symmetry and irrelevant
vortex operators.  In a gapless phase, the Luttinger parameter $K_p$
is defined by the general relation \cite{book}
\begin{equation}
\frac{\pi v}{K_p} = \frac{\partial ^2 E_p}{\partial N_p^2}
\label{luttdef}
\end{equation}
where $E_p$ is given in (\ref{Ep}), and $v$ is the velocity of
the excitations. Using the fact that $3N_t + 2N_p =N$ means that
$K_p = 9 K_t/4$, as noted above.

To do this computation, it is convenient to rewrite the Bethe equations
in terms of ``rapidity'' variables $\theta_j$. These
are defined by the relation
$$\omega_j \equiv e^{-3ik(\theta_j)}\equiv-\frac{\sinh(\theta_j +
i\gamma/2)} {\sinh(\theta_j - i\gamma/2)}$$ where $\gamma =
\cos^{-1}(-\Delta)$; for our models, $\Delta=1/3$.  This relation also
defines $k$ as a function of $\theta$, i.e. $k_j\equiv k(\theta_j)$.
This change of variables from momentum to rapidity is useful because
it puts the Bethe equations (\ref{bethe2}) in difference form: taking
the log of both sides gives
\begin{equation} (2M_j+1)\pi = RNk(\theta_j)   - \lambda k_{\rm tot} +
\sum_{m=1}^{N_p} L(\theta_j-\theta_m)
\label{bethediff}
\end{equation}
where $M_j$ is an integer, and
$$L(\theta)= i\ln \left[ \frac{\sinh(\theta + i\gamma)} {\sinh(\theta
- i\gamma)}\right].$$
We have defined the parameters $R$ and $\lambda$ to make the analysis
apply to general models with hard-core constraints. For the $p$ bosons
we have $R=1+ N_p/N$ and $\lambda=1$. For the $t$ bosons, we have
$R=1-N_t/N$ and $\lambda=-1$. The usual XXZ model has 
$R=1$ and $\lambda=0$. The cases discussed in
\cite{AB,FSN} can also be written in this form. 

Each kink is characterized by an different integer $M_j$; if one
were to have $M_j=M_m$ for $j\ne m$, then $\theta_j = \theta_m$,
and we showed above that identical bare momenta results in the
Bethe ansatz wavefunction vanishing. Summing (\ref{bethediff})
over $j$ and using the fact that $L(\theta)=-L(-\theta)$ means
that the total momentum is
\begin{equation}
k_{\rm tot} = \frac{1}{N}\sum_j (2M_j +1)\pi. \label{sumk}
\end{equation}
The
$\theta_j$ in the ground state are all real, and lie in the region
$|\theta_j|\le\Lambda$, where $\Lambda$ depends on the chemical potential
$h$.  There is one particle in the ground state
for each solution of (\ref{bethediff})
with $|\theta_j|\le\Lambda$; removing any particle in this region increases
the energy.

To make any more progress, one needs to
go to the thermodynamic limit, where there are many particles,
and the rapidities are closely spaced together.
We then define the density of particles $\rho(\theta)$
so that $N\rho(\theta)d\theta$ gives the number of
particles in the ground state
with rapidities between $\theta$ and $\theta+d\theta$.
In the ground state, there is a particle associated with
each integer $M_j$, as long as the rapidity is below $\Lambda$.
For $|\theta|<\Lambda$, the equations (\ref{bethediff}) therefore become
\begin{equation}
2\pi \rho(\theta) = Rk'(\theta) + \int_{-\Lambda}^\Lambda d\theta'
\ \Phi(\theta-\theta') \rho(\theta')
\label{dos}
\end{equation}
where $\Phi(\theta)\equiv L'(\theta)$.
The integral equation (\ref{dos}) determines the ground-state
density $\rho(\theta)$ for a particular $\Lambda$. The maximum
rapidity $\Lambda$ depends on $h_p$, and is determined by minimizing
the energy (\ref{Ep}), subject to
$\rho(\theta)$ obeying (\ref{dos}). In the thermodynamic
limit, the ground-state energy is
\begin{equation}
E_p = N\lint d\theta\ \rho(\theta) \epsilon_0(\theta)
\label{EpBA}
\end{equation}
where the bare energy of a kink is
$$\epsilon_0(\theta) = -h_p - \Delta - \frac{\sin^2(\gamma)}{\cosh(2\theta)
-\cos(\gamma)}.$$
The minimum value of the last term in $\epsilon_0$ is $-(1+\cos\gamma)=
\Delta-1$, so if $h_p \le -1 $,
it becomes energetically favorable to have
an empty ground state. Converting back to the original variables, this
is the Pokrovsky-Talapov transition occuring at $\sigma =-13/9$.

Even though the density $\rho(\theta)$ and the largest rapidity
$\Lambda$ completely characterize the system at zero temperature,
extracting the Luttinger parameter requires further work.
We first compute the velocity $v$ of the excitations. We define
$E_{\rm hole}(\theta_h)$ and $k_{\rm hole}(\theta_h)$ as the
energy and momentum change in the system resulting from removing a
particle of rapidity $\theta_h$ from the ground state. The
velocity is then
$$v= \frac{\partial E_{\rm hole}}{\partial
k_{\rm hole}}\Bigg|_{\theta_h=\Lambda}= \frac{\partial E_{\rm
hole}/\partial\theta_h}{\partial k_{\rm hole}/\partial\theta_h}
\Bigg|_{\theta_h=\Lambda}$$ Say we remove the particle associated
with integer $M_h$ and rapidity $\theta_h$.  Because the particles
are coupled, the momenta of all of them changes when a particle is
removed. However, the integers $M_j$ do not change, so from
(\ref{sumk}), we see that $k_{\rm
hole}=-(2M_h+1)\pi/N.$ Using (\ref{bethediff}), we
can rewrite this in the thermodynamic limit in terms of the
rapidities. After a few manipulations, we have
\begin{equation}
\frac{\partial k_{\rm hole}}{\partial\theta_h} = -2\pi
\rho(\theta_h) \label{khole}
\end{equation}
just as in the case without hard cores.

The energy of a hole of rapidity $\theta_h$ is
$$E_{\rm hole}(\theta_h) =
 -\epsilon_0(\theta_h) + \sum_{j\ne h} \epsilon_0'(\theta_j)
\delta\theta_j.$$
where $\delta\theta_j$ is
the change in rapidity of particle $j$ when the hole is created.
Because the integers $M_j$
do not change, (\ref{bethediff}) requires that
\begin{eqnarray*}
0&=& RNk'(\theta_j) \delta\theta_j + \lambda k(\theta_j) -\lambda
k_{\rm hole}
- L(\theta_j-\theta_h)\\
&&\qquad
+ \sum_{m\ne h} \Phi(\theta_j - \theta_m) (\delta\theta_j - \delta\theta_m).
\end{eqnarray*}
Taking the thermodynamic limit and
using (\ref{dos})  simplifies this to an integral equation:
\begin{eqnarray*}
2\pi N\rho(\theta)\delta\theta(\theta;\theta_h) &=& L(\theta-\theta_h)
-\lambda k(\theta) + \lambda k_{\rm hole}(\theta_h)\\ && -
N\lint d\theta' \Phi(\theta-\theta')
\rho(\theta')\delta\theta(\theta';\theta_h).
\end{eqnarray*}
Note that $\delta\theta$ is a function of both $\theta$ and $\theta_h$.
Let us define the function $\tau(\theta)$
as the solution of the integral equation
\begin{equation}
2\pi\tau(\theta) = -\Phi(\theta-\Lambda) + \lint d\theta' \Phi(\theta-\theta')
\tau(\theta').
\label{taudef}
\end{equation}
We then have
\begin{eqnarray*}
&&N\frac{\partial}{\partial \theta_h}\left[
\rho(\theta)\delta\theta(\theta;\theta_h) -
\rho(-\theta)\delta\theta(-\theta;\theta_h)\right]\Bigg|_{\theta_h =
\Lambda}\\
&&\qquad\qquad= \tau(\theta)-\tau(-\theta)
\end{eqnarray*}because $k_{\rm
hole}(\theta_h)$ is odd in $\theta_h$. Because
$\epsilon_0'(\theta)$ is odd in $\theta$, we use this with
expression for hole energy to get
\begin{equation}
\frac{\partial E_{\rm hole}}{\partial \theta_h}
\Bigg|_{\theta_h=\Lambda} = -\epsilon'_0(\Lambda) + \lint d\theta\
\epsilon_0'(\theta) \tau(\theta) \label{Ehole}
\end{equation}
As with $k_{\rm hole}$, this is the same as in the case without
hard cores. Combining this with (\ref{khole}) gives the velocity
of the excitations. We can check this by noting that as $\sigma\to
-13/9$, there are no $p$ bosons in the ground state. This means that
$\Lambda\to 0$, so it follows from (\ref{Ehole}), (\ref{khole})
and (\ref{dos}) that $v\to \epsilon'_0(0)/k'(0) = 0$. The
quasiparticles here have dispersion $E_{\rm hole}\propto k_{\rm
hole}^2$, as expected at a $z=2$ Pokrovsky-Talapov transition.

The computation of the
Luttinger parameter (\ref{luttdef}) is affected by the hard-exclusion effects.
In particular, because $R=(1+N_p/N)/3$ in (\ref{dos}) depends on
the number of particles $N_p$, $\partial R/\partial\Lambda \ne 0$.
First we compute how the number of particles changes when $\Lambda$ is varied.
It is convenient to define the rescaled density $\trho = \rho/R$,
so that for fixed $\Lambda$, $\trho$ is independent of $R$. Then if we
define $\chi(\theta)$ as the solution of the integral equation
\begin{equation}
2\pi \chi(\theta) = L(\theta-\Lambda) + L(\theta+\Lambda)
+\lint d\theta' \Phi(\theta-\theta') \chi(\theta')
\label{chidef}
\end{equation}
it is straightforward to show that
$$\frac{\partial\trho(\theta)}{\partial\Lambda}=\frac{\trho(\Lambda)}
{1-\chi(\Lambda)}\frac{\partial \chi(\theta)}{\partial\theta}.$$
Because
$$N \lint d\theta\, \trho(\theta) =
\frac{N_p}{R} = \frac{N_p}{1+N_p/N},$$
we have
$$ N_p'\equiv \frac{\partial N_p}{\partial\Lambda}= N {R^2}
\frac{2\trho(\Lambda)}{1-\chi(\Lambda)}.$$

To compute the Luttinger parameter, it is convenient to also define a rescaled
energy $\widetilde E=E_p/R$. A little algebra then yields
$$\frac{\partial ^2 E_p}{\partial N_p^2} =  \frac{1}{RN_p'}
\frac{\partial}{\partial \Lambda}\left( \frac{\widetilde E' R^2}{N_p'}\right)$$
where $\widetilde E' \equiv \partial \widetilde E /\partial \Lambda$.
Computing $\widetilde E'$ is
then similar to the $R=1$ case of \cite{haldanelutt}, yielding
$$\frac{\partial ^2 E_p}{\partial N_p^2} =  \frac{1}{RN_p'}
\frac{\partial}{\partial \Lambda}
\left(\epsilon_0(\theta) - \frac{1}{2}
\lint \epsilon_0'(\theta) \chi(\theta)\right).$$
Since none of the quantities inside the derivative depends on $R$,
we can use the calculation of \cite{haldanelutt} to get
$$\frac{\partial ^2 E_p}{\partial N_p^2} =
\frac{1}{R^3}
\frac{(1-\chi(\Lambda))^2}{2\trho(\Lambda)}
\left(\epsilon_0'(\theta) - \lint \epsilon_0'(\theta) \tau(\theta)\right)
$$
where $\tau(\theta)$ is defined above via (\ref{taudef}).
Combining this with our computation of the velocity above yields finally the
Luttinger parameter
\begin{equation}
K_p= \left(\frac{1+N_p/N}{1-\chi(\Lambda)}\right)^2
\label{luttparam}
\end{equation}
The effect of the hard-core constraints ends up being the $(1+N_p/N)^2$
in the numerator.

We have seen that when the coupling
$\sigma = -13/9$, there are no $p$ bosons in the ground state.
The equation (\ref{luttparam}) gives the correct $K_p=1$,
because $\chi(0)=0$ and $N_p=0$ here.
As $\sigma$ is
increased, $p$ bosons with $|\theta|\le \Lambda$ fill the ground state. For
some value of $\sigma$, $\Lambda\to \infty$. We can find out how many
bosons there are here, because in this limit we can solve
the equation (\ref{dos}) by Fourier transformation. Actually, to just
get $N_p(\Lambda\to\infty)$, we do not need to go to the trouble: we know
from the XXZ model that
$$\lim_{\Lambda\to\infty} \lint d\theta\, \trho(\theta)=\frac{1}{2}.$$
Thus in our case, we have $N_p(\Lambda\to\infty)= N/5$.  The equation
(\ref{luttparam}) for the Luttinger parameter is therefore valid only
for values of chemical potential such that only solutions of the Bethe
equations with all bare momenta real are present in the ground state.
In other words, (\ref{luttparam}) is valid only for
$\sigma$ such that $N_p \le N/5$.
To find the Luttinger parameter for the remaining range of $\sigma$, we
can repeat the above analysis for the $t$ bosons. This yields
\begin{equation}
K_t= \left(\frac{1-N_t/N}{1-\chi(\Lambda_t)}\right)^2,
\label{luttparam2}
\end{equation}
where $\Lambda_t$ is the maximum rapidity of a $t$ boson in the ground state.
By the same argument, this expression is valid for $N_t\le N/5$.

Because $2N_p+3N_t=N$, the expressions for the Luttinger parameter
are both valid at exactly one point: where $\Lambda=\Lambda_t =
\infty$, $N_p=N_t=1/5$.  This provides a good check on our result: at
$\Lambda=\Lambda_t=\infty$, $N_p=N_t=1/5$, comparing the
expressions (\ref{luttparam}) and (\ref{luttparam2}) yields $K_t =
4 K_p /9$, as noted in general in (\ref{K1}). In fact, we can find
the exact value of $K_p$ at this point, by either a Weiner-Hopf
analysis \cite{BIK}, or by appealing to the XXZ model, where the
answer is known. Either way, one finds that
$$\lim_{\Lambda\to\infty} (1-\chi(\Lambda))^2 = 2\cos^{-1}(\Delta)/\pi$$
For our case of $\Delta=1/3$, we have therefore
$K_p = 1.83754$, so vortices are relevant here.
Another check on our results is to note that the computation of
the velocity $v$ must give the same answer at this point whether computing
with the $t$ bosons or the $p$ bosons. Indeed, we see from (\ref{khole})
and (\ref{Ehole}) that both are independent of $R$ and $\lambda$, only
depending on $\Lambda$; since $\Lambda$ is the same for both pictures,
we do get the same answer.

\begin{figure}
\centerline{\includegraphics[width=3.5in]{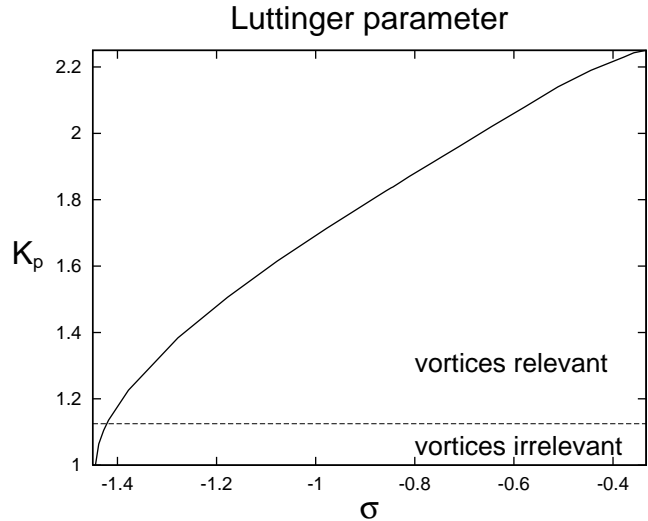}}
\caption{Bethe ansatz results for $K_p$ as a function of $\sigma$.
Recall that $K_t = 4 K_p /9$.} \label{luttplot}
\end{figure}

The incommensurate regime is characterized by having $K_p<9/8$. We
cannot analytically find the value of $\sigma$ where $K_p = 9/8$, but
it is straightforward to solve the above integral equations
numerically. First, one must determine the value of $\Lambda$ for a
given $\sigma$ by solving (\ref{dos}) for $\rho(\theta)$ and $\Lambda$
while demanding that the energy (\ref{EpBA}) be minimized. Integrating
$\rho(\theta)$ yields the ground-state particle density $N_p/N$ as
well.  Knowing $\Lambda$ then allows one to solve (\ref{chidef}) for
$\chi(\theta)$ and hence $\chi(\Lambda)$. Plugging this into
(\ref{luttparam}) then yields the Luttinger parameter. We find that
$K_p=9/8$ when $\sigma =1.422...$. This leads to our final result for
the phase diagram: the incommensurate phase exists only for
$$ 1.422 < \sigma < 1.444\ . $$
The incommensurate region is also very
narrow near the Potts critical point $M_3$, so it is likely that it
occupies a very small region of parameter space.
We plot $K_p$ as obtained above in
Fig~\ref{luttplot}.

\section{Conclusion}

The primary purpose of this paper was to provide a thorough study
of the very simple one-dimensional boson model in (\ref{ham}).
This model was originally motivated by the studies of `tilted'
Mott insulators of atoms in optical lattices. Despite its
simplicity, our model has a rich phase diagram, shown in
Fig~\ref{phase}, with a many competing ordered phases and
possibilities for exact solutions.

Our solutions shed additional light on the nature of the phase
transition in the chiral clock model. Over a certain regime of
parameters, we provide definitive evidence for a gapless, floating
incommensurate phase adjacent to the gapped ${\bf Z}_3$ ordered phase
with a period-3 density wave. However, the size of this
incommensurate phase was remarkably small, making it unlikely that
it will ever be observed in experimental or numerical studies. We
also obtained specific realizations of two of the exactly soluble
multi-critical points identified by Andrews {\em et al.}
\cite{abf}, which are now known to be part of the series of
minimal conformal models \cite{huse}. It is likely that our
analysis could be generalized to this series; it would be interesting
in particular to see if the incommensurate phases existed in general.

\begin{acknowledgments}
We are very grateful to David Huse and Nick Read for a number of
useful comments.  This research was supported by US NSF grants DMR-0098226
(K.S. and S.S.) DMR-0104799 (P.F.), and by a DOE grant
DEFG02-97ER41027 (P.F.). P.F.\  and S.S.\ would also like to thank the
Aspen Center for Physics for hospitality during part of this work.

\end{acknowledgments}

\appendix*

\section{Connection to Baxter's hard square model}

In this appendix we show that the quantum Hamiltonian
$\mathcal{H}$ in (\ref{h}) generates precisely the transfer matrix
of the classical, two-dimensional hard-square model introduced by
Baxter \cite{square}.

It is useful to note that the hard-square model, as defined in the
Introduction, can also be interpreted as a
certain `interface' or `height' model. In particular, it is
identical to the so-called $A_4$ restricted-solid-on-solid (RSOS)
\cite{abf}. To
see the connection with the RSOS model, we need to associate each
tile configuration with a corresponding set of heights. On one
sublattice of the square lattice,
associate each site with a tile present with height
$1$, and a vacancy with height $3$. On the other sublattice,
associate each tile with height $4$, and a vacancy with height
$2$. Then the hard-square restrictions are simply encoded in a set
of constraints on heights on nearest neighbor sites: height $1$
can only be next to $2$, $2$ can be next to $3$ or $1$, $3$ can
only be next to $2$ or $4$, and $4$ can only be next to $3$.
These height restrictions are conveniently summarized
in the $A_4$ Dynkin diagram, and hence the terminology.
Like the hard-square case,
the general $A_n$ RSOS models are solvable along two lines, with one
critical point on each. These critical points are multicritical points
with known conformal-field-theory descriptions \cite{huse}.

\begin{figure}[ht]
\begin{picture}(210,30)
\put(0,0){\line(1,1){15}} \put(15,15){\line(1,-1){15}}
\put(30,0){\line(1,1){15}} \put(45,15){\line(1,-1){15}}
\put(60,0){\line(1,1){15}} \put(75,15){\line(1,-1){15}}
\put(90,0){\line(1,1){15}} \put(105,15){\line(1,-1){15}}
\put(120,0){\line(1,1){15}} \put(135,15){\line(1,-1){15}}
\put(150,0){\line(1,1){15}} \put(165,15){\line(1,-1){15}}
\put(180,0){\line(1,1){15}} \put(195,15){\line(1,-1){15}}
\end{picture}
\caption{Zig-zag line defining the space of states upon which
$\mathcal{H}$ acts.} \label{wt2}
\end{figure}
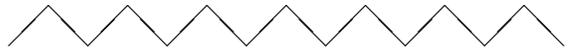

Here, we continue to use the terminology of hard squares. We
consider the diagonal-to-diagonal transfer matrix of the hard
square model. In the orientation of Fig.~\ref{wt1}, this transfer
matrix acts on the space of states of a zig-zag line of lattice
sites as shown in Fig.~\ref{wt2}.

There are three interaction parameters: $L$ and $M$ are the
interaction strength of adjacent tiles, while $z$ is the fugacity
for each tile. The grand canonical partition function can then be
written as the sum of products of Boltzmann weights. The Boltzmann
weight of a plaquette is shown in Fig.~\ref{wt1}
\begin{figure}[ht]
\begin{picture}(230,50)
\put(30,0){$a$} \put(7,23){$b$} \put(30,10){\line(-1,1){15}}
\put(30,10){\line(1,1){15}} \put(51,23){$d$}
\put(45,25){\line(-1,1){15}} \put(15,25){\line(1,1){15}}
\put(30,45){$c$} \put(80,22){$= \ z^{(a+b+c+d)/4} e^{Lac+Mbd}
t^{-a+b-c+d}$}
\end{picture}
\caption{Boltzmann weight of a plaquette of the square lattice. The
tiles are centered on the vertices of this plaquette, and $a,b,c,d
= 0,1$ indicates absence/presence of a tile.} \label{wt1}
\end{figure}
where the heights $a,b,c$ and $d$ are $0$ or $1$; $1$ denotes a
tile, so the restriction is that $ab=bc=cd=da=0$. The extra
parameter $t$ cancels out of the partition function and so is
arbitrary. The pure hard square model is obtained by setting
$L=M=0$, but we will need the full range of values of these
couplings to explore the parameter space of $\mathcal{H}$. The
hard-hexagon model (on a triangular lattice) is obtained by
setting $L=0$ and $M\to-\infty$.

We will now take the so-called $\tau$-continuum limit of this
transfer matrix, so that we may obtain the corresponding
Hamiltonian which evolves the states in a continuous imaginary
time. To take this Hamiltonian limit, we define $t=\exp(L/4)$ and
$z=e^{-L}(1-\zeta)$. Then the Boltzmann weights for all the
allowed possible states around each plaquette are shown in
Fig.~\ref{wt3}.
\begin{figure}
\begin{picture}(100,60)
\put(28.5,0){$0$} \put(7,23){$1$} \put(30,10){\line(-1,1){15}}
\put(30,10){\line(1,1){15}} \put(51,23){$1$}
\put(45,25){\line(-1,1){15}} \put(15,25){\line(1,1){15}}
\put(28.5,45){$0$} \put(80,22){$= \ (1-\zeta)^{1/2}e^{M}$}
\end{picture}\\
\begin{picture}(100,60)
\put(28.5,0){$1$} \put(7,23){$0$} \put(30,10){\line(-1,1){15}}
\put(30,10){\line(1,1){15}} \put(51,23){$0$}
\put(45,25){\line(-1,1){15}} \put(15,25){\line(1,1){15}}
\put(28.5,45){$1$} \put(80,22){$= \ (1-\zeta)^{1/2}$}
\end{picture}\\
\begin{picture}(100,60)
\put(28.5,0){$0$} \put(7,23){$0$} \put(30,10){\line(-1,1){15}}
\put(30,10){\line(1,1){15}} \put(51,23){$1$}
\put(45,25){\line(-1,1){15}} \put(15,25){\line(1,1){15}}
\put(28.5,45){$0$} \put(80,22){$= \ (1-\zeta)^{1/4}$}
\end{picture}\\
\begin{picture}(100,60)
\put(28.5,0){$1$} \put(7,23){$0$} \put(30,10){\line(-1,1){15}}
\put(30,10){\line(1,1){15}} \put(51,23){$0$}
\put(45,25){\line(-1,1){15}} \put(15,25){\line(1,1){15}}
\put(28.5,45){$0$} \put(80,22){$= \ (1-\zeta)^{1/4}e^{-L/2}$}
\end{picture}\\
\begin{picture}(100,60)
\put(28.5,0){$0$} \put(7,23){$0$} \put(30,10){\line(-1,1){15}}
\put(30,10){\line(1,1){15}} \put(51,23){$0$}
\put(45,25){\line(-1,1){15}} \put(15,25){\line(1,1){15}}
\put(28.5,45){$0$} \put(80,22){$= \ 1$}
\end{picture}
\caption{Boltzmann weights for the possible states of a plaquette
of the hard square model} \label{wt3}
\end{figure}

When $\zeta=M=0$ and $L\to\infty$, the vertical transfer matrix is
the identity. The Hamiltonian is found by expanding around this
limit. The small parameters are then $\zeta$, $M$ and $e^{-L/2}$.
Note that to get back to the original Hilbert space, we need to
break the transfer matrix into two pieces. The first evolves the
lower of the two rows of sites in the above zig-pattern, and then
the second evolves the upper row. The full transfer matrix is then
\begin{equation}
T_1T_2\approx (1- \zeta \mathcal{H}_1)(1- \zeta \mathcal{H}_2)
\approx (1-\zeta \mathcal{H}),
\end{equation}
where $\mathcal{H}=\mathcal{H}_1+\mathcal{H}_2$. Let $d_j^\dagger$
be a boson creation operator which creates a tile on site $j$. We
identify these tiles with the bosons of (\ref{h}). It is easy to
see that the constraints of the hard square model correspond
precisely to the constraints (\ref{constraints}). Moreover, here
we obtain the same Hamiltonian in (\ref{h}) (up to an arbitrary
overall energy scale) with the couplings
\begin{equation}
\frac{U}{w} = e^{L/2} \zeta~~~;~~~ \frac{V}{w} = -M e^{L/2}
\end{equation}
Note that $M>0$ means attractive next-nearest-neighbor
interactions.

Baxter \cite{square} showed that the hard-square model
is integrable for
\begin{equation}
z=\frac{(1-e^{-L})(1-e^{-M})}{e^{L+M}-e^L-e^M}. \label{abf1}
\end{equation}
In the Hamiltonian limit, this condition is equivalent to
(\ref{intlines}). Baxter also showed
that there are two critical points within the parameter space of
(\ref{abf1}). Defining
$$I=z^{-1/2}(1-ze^{L+M}),$$
the critical points are at
$$I= \pm \left[\frac{1+\sqrt{5}}{2}\right]^{-5/2},$$
which leads to (\ref{so},\ref{crit}). The critical point at (\ref{so}) is
in the universality class of the tricritical Ising model, while
that at (\ref{crit}) is of the 3-state Potts type \cite{square,husesquare}.
The physical interpretation of these critical points is discussed
in Section~\ref{sec:phase}.

\end{document}